\documentclass{aa}
\usepackage{graphicx}
\usepackage{longtable}
\usepackage[varg]{txfonts}
\begin{document}
   \title{Photometric and spectroscopic variability of the FUor star V582 Aurigae}


   \author{E. H. Semkov\inst{1}
          \and    
          S. P. Peneva\inst{1}
          \and
          U. Munari\inst{2}
          \and
          M. Dennefeld\inst{3}
          \and
          H. Mito\inst{4}
          \and
          D. P. Dimitrov\inst{1}
          \and
          S. Ibryamov\inst{1}
          \and
          K. A. Stoyanov\inst{1}}

   \offprints{E. Semkov (esemkov@astro.bas.bg)}

   \institute{Institute of Astronomy and National Astronomical Observatory, Bulgarian Academy of Sciences,
              72 Tsarigradsko Shose blvd., BG-1784 Sofia, Bulgaria,
              \email{esemkov@astro.bas.bg}
         \and
         INAF Osservatorio Astronomico di Padova, Sede di Asiago, I-36032 Asiago (VI), Italy
         \and
         Institut d'Astrophysique de Paris, CNRS, and Universit\'{e} Pierre et Marie Curie, 98 bis Boulevard Arago, 75014 Paris, France
         \and
         Kiso Observatory, Institute of Astronomy, School of Science, the University of Tokyo, Mitake-mura, Kiso-gun, Nagano-ken 397-0101, Japan
         }

   \date{Received ; accepted }


  \abstract
   {}
   {We present results from optical photometric and spectroscopic observations of the eruptive pre-main sequence star V582 Aur. 
   Variability of the star was reported a few years ago when it was suspected as a possible FU Orionis object.
   Due to the small number of currently known FUors, a new object of this type is ideal target for follow-up photometric and spectroscopic observations.}
   {We carried out $BVRI$ CCD photometric observations in the field of V582 Aur from 2009 August to 2013 February. 
   We acquired high-, medium-, and low-resolution spectroscopy of V582 Aur during this period. 
   To study the pre-outburst variability of the target and construct its historical light curve, we searched for archival observations in photographic plate collections. 
   Both CCD and photographic observations were analyzed using a sequence of 14 stars in the field of V582 Aur calibrated in $BVRI$.}
   {The pre-outburst photographic observations of V582 Aur show low-amplitude light variations typical of T Tauri stars. 
   Archival photographic observations indicate that the increase in brightness began in late 1984 or early 1985 and the star reached the maximum level of brightness at 1986 January.
   The spectral type of V582 Aur can be defined as G0I with strong P Cyg profiles of H$\alpha$ and Na I D lines, which are typical of FU Orionis objects.
   Our $BVRI$ photometric observations show large amplitude variations $\Delta V\sim2\fm8$) during the 3.5 year period of observations. 
   Most of the time, however, the star remains in a state close to the maximum brightness. 
   The deepest drop in brightness was observed in the spring of 2012, when the brightness of the star fell to a level close to the pre-outburst.
   The multicolor photometric data show a color reversal during the minimum in brightness, which is typical of UX Ori variables.
   The corresponding spectral observations show strong variability in the profiles and intensities of the spectral lines (especially H$\alpha$), which indicate significant changes in the accretion rate.
   On the basis of photometric monitoring performed over the past three years, the spectral properties of the maximal light, and the shape of the long-term light curve, we confirm the affiliation of V582 Aur to the group of FU Orionis objects.}
   {}


   \titlerunning{Photometric and spectroscopic variability of V582 Aurigae}
   \maketitle
%

\section{Introduction}
Photometric and spectral variability is the most common characteristic of pre-main sequence (PMS) stars.
Both classes of PMS stars, the wide-spread low mass ($\it M$ $\leq$ $2M_{\sun}$) T Tauri stars and the more massive Herbig Ae/Be stars, show various types of variability.
The main physical mechanisms causing the brightness variation of PMS stars are defined by Herbst et al. (1994, 2007) and Bouvier et al. (1995).
The study of the large amplitude brightness variations of PMS stars is of great importance in understanding stellar evolution. 
These variations comprise transient increases in brightness (outbursts), temporary drops in brightness (eclipses), and large amplitude irregular or regular variations for a short or long time scales.
In many cases large amplitude variations in brightness are accompanied by changes of the spectral type or by variability in the profiles and the presence of individual spectral lines (Herbig et al. 2003, Grinin et al. 2001, Rodgers et al. 2002, Kurosawa \& Romanova 2013).
This is especially true for the H$\alpha$ emission line, which is the most prominent feature in the PMS spectra.

The large amplitude outbursts of PMS stars can be grouped into two main types, named after their respective prototypes: FU Orionis (FUor; Ambartsumian 1971) and EX Lupi (EXor; Herbig 1989).
The flare-up of FU Orionis itself occurred in 1936 and for several decades it was the only known object of that type.
Herbig (1977) defined FUors as a class of young variables after the discovery of two new FUor objects, V1057 Cyg and V1515 Cyg.
Several more objects were assigned to this class of young variables over the next four decades (see Reipurth \& Aspin 2010 and references therein).

Because only a small number of FUor stars have been detected to date, photometric and spectral studies of every new object are of great interest.
Due to the large-scale optical and infrared monitoring programs carried out in several observatories and the contributions of amateur astronomers, some new objects have been observed to undergo large amplitude outbursts, V733 Cep (Reipurth et al. 2007, Peneva et al. 2010), V2493 Cyg (Semkov et al. 2010, 2012; Miller et al. 2011, K{\'o}sp{\'a}l et al. 2011), V2492 Cyg (Aspin 2011, Hillenbrand et al. 2013,  K{\'o}sp{\'a}l et al. 2013), V2494 Cyg (Aspin et al. 2009a), V2495 Cyg (Movsessian et al. 2006), V900 Mon (Reipurth et al. 2012), V2775 Ori (Fischer et al. 2012).

A typical outburst of FUor objects can last for several decades, and the rise time is shorter than that of the decline.
All known FUors share the same defining characteristics: a $\Delta$$V$$\approx$4-6 mag. outburst amplitude, association with reflection nebulae, location in star-forming regions, an F-G supergiant spectrum during outbursts, a strong LiI~6707~\AA\ line in absorption, and CO bands in near-infrared spectra (Herbig 1977, Reipurth \& Aspin 2010).  
Typically the decrease in brightness goes smoothly, but several events of temporary drops in brightness have been registered in the cases of V1515 Cyg (Kenyon et al. 1991, Clark et al. 2005), V733 Cep (Peneva et al. 2010) and V1735 Cyg (Peneva et al. 2009).
An important feature of FUors is the massive supersonic wind observed as a P Cyg profile most commonly for both H$\alpha$ and Na I D lines.

The EXor objects undergo frequent, irregular, and relatively brief (a few weeks to a few months or one year) outbursts with amplitudes $\Delta$$V$$\approx$2-5 mag. 
During such events, the cool spectrum of the quiescence is veiled, and strong emission lines from single ionized metals are observed together with appearance of reversed P-Cyg absorption components (Herbig 2007).

Both types of eruptive stars, FUors and EXors, seem to be related to the low-mass PMS objects (T Tauri stars), which have massive circumstellar disks. 
These objects have been classified in terms of their wide range of available photometric and spectral properties, but their outbursts are thought to have the same cause: an enhanced accretion rate from the circumstellar disk onto the central star (Hartmann \& Kenyon 1996, Herbig 2007).
At the time of their outbursts, FUor objects undergo an increase in their accretion rate from $\sim$10$^{-7}$$M_{\sun}$$/$yr up to $\sim$10$^{-4}$$M_{\sun}$$/$yr.
The periods of enhanced accretion are thought to be triggered by thermal or gravitational instability in the circumstellar disk (Hartmann \& Kenyon 1996, Zhu et al. 2009).
Another possible triggering mechanism could be the interactions of the circumstellar disk with a planet or nearby stellar companion on an eccentric orbit (Lodato \& Clarke 2004, Reipurth \& Aspin 2004, Pfalzner 2008).
For a period of $\sim$ 100 years, the circumstellar disk adds $\sim10^{-2}$ M$\sun$ onto the central star and ejects $\sim$10\% of the accreting material in a high-velocity stellar wind. 
Some FUor objects were found to exhibit periodic spectroscopic (Herbig et al. 2003, Powell et al. 2013) or low amplitude photometric (Kenyon et al. 2000, Green et al. 2013, Siwak et al. 2013) variability in short time-scale (days).

Approximately 25$\%$ of Herbig Ae/Be stars and some T Tauri stars of F-G type show strong photometric variability with sudden quasi-Algol drops of brightness and amplitudes up to 3 mag. ($V$) (Natta et al. 1997).
During the deep minima of brightness, an increase of polarization and specific color variability are observed.
The prototype of this group of PMS objects with intermediate mass named UXors is UX Ori.
The widespread explanation of its variability are variable obscurations from orbiting circumstellar clumps of dust or edge-on circumstellar disks (Grinin et al. 1991).

The discovery of a new FUor candidate in Auriga was reported by Anton Khruslov (Samus 2009). 
The star was cataloged as USNO A2.0 1200-03303169 and according to the General Catalog of Variable Stars as V582 Aur. 
V582 Aur is located in a region of active star formation near Auriga OB2 association (Fig. 1).
Samus (2009) examined the brightness of the star on photographic plates from the Moscow collection (1965-1992) and on the images from the Digitized Sky Survey (1954-1993), which suggests that the increase in brightness started between 1982 and 1986. 
Munari et al. (2009) obtained the first low-resolution spectrum of V582 Aur on 2009 August 6 and registered the presence of absorption lines of the Balmer series, Na I D and Ba II ($\lambda$ 6496) and the absence of the Li I ($\lambda$ 6707) line in the spectrum. 
The photometric observations of V582 Aur reported by Munari et al. (2009) show the star near to maximal brightness on 2009 August 5 and 6.
Using the mid-infrared part of the AKARI All-Sky Survey, Takita et al. (2010) identified V582 Aur as a T Tauri star candidate.
Our first results from photometric and spectral observations of the star were reported in Semkov et al. (2011). 
In the paper we come to the conclusion that V582 Aur has all observational characteristics of FUor objects.

\begin{figure}
   \centering
   \includegraphics[width=8cm]{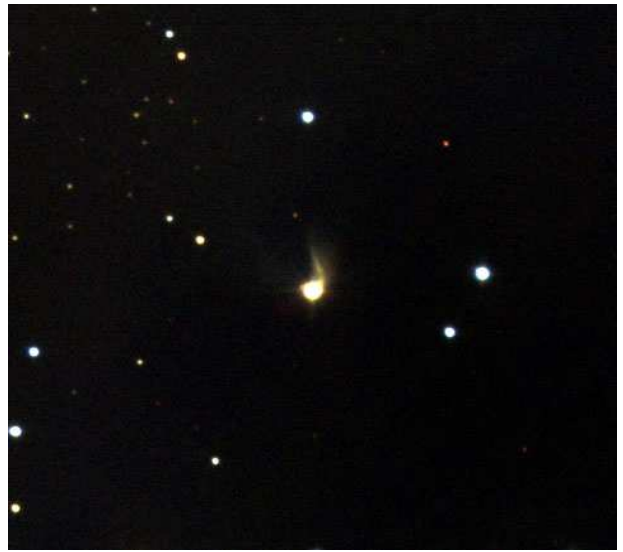}
   \caption{Color image of V582 Aur obtained with the 2 m RCC telescope in NAO Rozhen. A faint cometary nebula arising from the star is clearly visible.}
    \end{figure}

Recent data from photometric and spectral observations and results from archival photographic plate measurements are reported in the present paper. 
We try to collect regular observations (spectroscopy and multicolor photometry) of V582 Aur in order to clarify the nature of variability of the object.

\section{Observations}

\subsection{Photometric CCD observations}
 
The CCD photometric observations of V582 Aur were performed with the 2 m RCC, the 50/70 cm Schmidt, and the 60 cm Cassegrain telescopes of the National Astronomical Observatory (NAO) Rozhen (Bulgaria) and with the 1.3 m RC telescope of the Skinakas Observatory\footnote{Skinakas Observatory is a collaborative project of the University of Crete, the Foundation for Research and Technology - Hellas, and the Max-Planck-Institut f\"{u}r Extraterrestrische Physik.} of the Institute of Astronomy, University of Crete (Greece).  
Observations were performed with four types of the CCD camera $—$ Vers Array 1300B at the 2 m RCC telescope, ANDOR DZ436-BV at the 1.3 m RC telescope, FLI PL16803 at the 50/70 cm Schmidt telescope, and FLI PL9000 at the 60 cm Cassegrain telescope.
All frames were exposed through a set of standard Johnson-Cousins filters.
All the data were analyzed using the same aperture, which was chosen as 4\arcsec in radius (while the background annulus was from 9\arcsec to 14\arcsec) in order to minimize the light from the surrounding nebula.

In order to facilitate transformation from instrumental measurement to the standard Johnson-Cousins system, fourteen stars in the field of V582 Aur were calibrated in $BVRI$ bands. 
Calibration was made during seven clear nights in 2010 and 2011 with the 1.3 m RC telescope of the Skinakas Observatory.
Standard stars from Landolt (1992) were used as a reference.  
Table 1 contains photometric data for the $BVRI$ comparison sequence. 
The corresponding mean errors in the mean are also listed. 
The stars are labeled from A to N in order of their V-band magnitude.  
In regions of star formation a great percentage of stars can be photometric variables.  
Therefore, there is a possibility that some of our standard stars are low amplitude variables, and we advise observers to use our photometric sequence with care.  
The finding chart of the comparison sequence is presented in Fig. 2.  
The field is $8\arcmin\times8\arcmin$, north is at the top and east is to the left.  
The chart is retrieved from the STScI Digitized Sky Survey Second Generation Red.

\begin{table*}
\caption{Photometric data for the $BVRI$ comparison sequence.}
\label{table:1}
\centering
\begin{tabular}{ccccccccc}
\hline\hline
\noalign{\smallskip}
Star &  $B$ & $\sigma_B$ & $V$ & $\sigma_V$ & $Rc$ & $\sigma_R$& $Ic$ & $\sigma_I$ \\
\noalign{\smallskip}
\hline
\noalign{\smallskip}
A & 15.635 & 0.037 & 14.153 & 0.023 & 13.303 & 0.055 & 12.486 & 0.013\\
B & 16.120 & 0.053 & 14.303 & 0.026 & 13.242 & 0.063 & 12.144 & 0.013\\
C & 16.271 & 0.028 & 15.324 & 0.023 & 14.766 & 0.041 & 14.156 & 0.017\\
D & 16.708 & 0.044 & 15.774 & 0.024 & 15.193 & 0.039 & 14.529 & 0.020\\
E & 17.254 & 0.049 & 16.206 & 0.028 & 15.590 & 0.046 & 14.982 & 0.018\\
F & 17.148 & 0.073 & 16.250 & 0.021 & 15.727 & 0.036 & 15.161 & 0.029\\
G & 17.382 & 0.063 & 16.267 & 0.026 & 15.620 & 0.046 & 14.920 & 0.024\\
H & 17.298 & 0.061 & 16.282 & 0.022 & 15.682 & 0.045 & 14.930 & 0.020\\
I & 18.055 & 0.153 & 16.584 & 0.044 & 15.672 & 0.060 & 14.748 & 0.014\\
J & 17.706 & 0.051 & 16.779 & 0.034 & 16.190 & 0.049 & 15.599 & 0.048\\
K & 18.642 & 0.099 & 17.004 & 0.029 & 15.990 & 0.064 & 14.839 & 0.030\\
L & 17.948 & 0.110 & 17.016 & 0.042 & 16.531 & 0.030 & 15.816 & 0.037\\
M & 18.276 & 0.156 & 17.154 & 0.054 & 16.469 & 0.046 & 15.714 & 0.034\\
N & 18.418 & 0.131 & 17.301 & 0.029 & 16.656 & 0.047 & 15.944 & 0.049\\
\hline                                   
\end{tabular}
\end{table*}

\begin{figure}
   \centering
   \includegraphics[width=8cm]{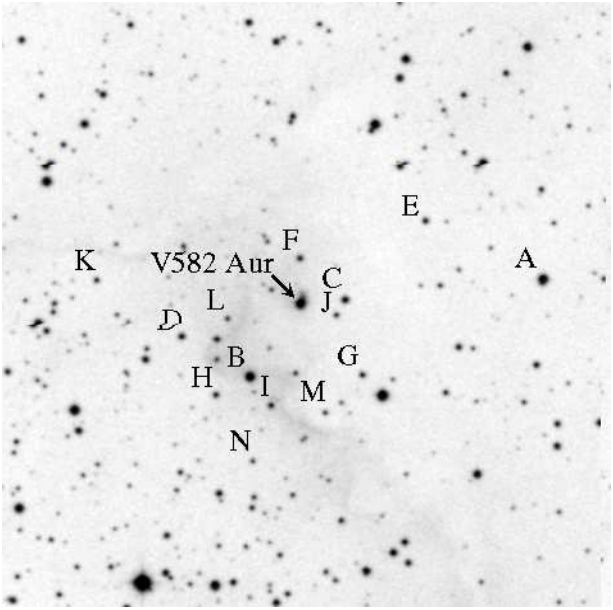}
   \caption{Finding chart for the $\it BVRI$ comparison sequence around V582 Aur.}
    \end{figure}

The results from our photometric CCD observations of V582 Aur are summarized in Table 2.  
The measured magnitudes from the photometric observations of V582 Aur made with the Asiago Schmidt telescope on 2009 August are also included in the table.
All CCD frames are measured with the same parameters and standard stars reported in the present paper.
The columns of Table 2 provide the date and Julian date (J.D.) of observation, $\it IRVB $ magnitudes of V582 Aur, and the telescope and CCD camera used. 
The typical errors in the reported magnitudes are $0\fm01$-$0\fm02$ for $I$ and $R$-band data, $0\fm01$-$0\fm03$ for $V$, and $0\fm02-0\fm04$ for $B$-band.

\onllongtab{2}{
\begin{longtable}{llllllll}
\caption{Photometric CCD observations of V582 Aur during the period 2009-2013.}\\
\hline
\hline
\noalign{\smallskip}
Date & J.D. (24...) & I & R & V & B & Telescope & CCD	\\
\noalign{\smallskip}
\hline
\endfirsthead
\caption{Continued.}\\ 
\hline
\hline
\noalign{\smallskip}
Date & J.D.(24...) & I & R & V & B & Telescope & CCD	\\
\noalign{\smallskip}
\hline
\endhead
\hline
\endfoot
\hline
\noalign{\smallskip}
\endlastfoot
\noalign{\smallskip}
2009 Aug 05 & 55048.581 & 12.26 & 13.60 & 14.85 & 16.72 & Asiago Sch  & \\
2009 Aug 06 & 55049.592 & 12.21 & 13.59 & 14.91 & 16.62 & Asiago Sch  & \\
2009 Oct 07 & 55111.519 & 12.70 & -     & -     & -     & Sch  & FLI\\						
2009 Oct 07 & 55112.491 & 12.70 & 14.29 & -     & -     & Sch  & FLI\\
2009 Oct 08 & 55113.491 & 12.67 & 14.24 & -     & -     & Sch  & FLI\\
2009 Oct 09 & 55114.491 & 12.66 & 14.24 & 15.53 & -     & Sch  & FLI\\
2009 Oct 28 & 55133.397 & 12.49 & 14.03 & 15.27 & -     & Sch  & FLI\\
2009 Nov 19 & 55155.352 & 12.34 & 13.84 & 15.06 & 16.83 & Sch  & FLI\\
2009 Nov 20 & 55156.418 & 12.33 & 13.83 & 15.05 & 16.80 & Sch  & FLI\\
2009 Nov 21 & 55157.449 & 12.34 & 13.83 & 15.06 & 16.81 & Sch  & FLI\\
2009 Nov 25 & 55161.476 & 12.28 & 13.72 & 14.94 & 16.73 & 2m   & VA\\
2010 Mar 11 & 55267.327 & 12.34 & 13.80 & 15.01 & -     & 2m   & VA\\
2010 Mar 12 & 55268.247 & 12.32 & 13.79 & 15.00 & 16.70 & 2m   & VA\\
2010 Apr 13 & 55300.294 & 12.11 & -     & -     & -     & Sch  & FLI\\
2010 May 14 & 55331.281 & 12.53 & 14.02 & -     & -     & Sch  & FLI\\
2010 May 16 & 55333.286 & 12.43 & 13.96 & 15.30 & -     & Sch  & FLI\\
2010 Aug 06 & 55415.577 & 12.60 & 14.13 & 15.38 & 17.04 & Sch  & FLI\\
2010 Aug 07 & 55416.549 & 12.60 & 14.14 & 15.40 & 17.17 & Sch  & FLI\\
2010 Aug 12	&	55420.572	&	12.63	&	14.17	&	15.45	&	17.20	&	1.3m & ANDOR\\
2010 Aug 13	&	55421.566	&	12.65	&	14.19	&	15.47	&	17.22	&	1.3m & ANDOR\\
2010 Aug 15	&	55423.598	&	12.71	&	14.29	&	15.60	&	-     &	1.3m & ANDOR\\
2010 Aug 24	&	55432.538	&	12.11	&	13.50	&	14.66	&	16.29	&	1.3m & ANDOR\\
2010 Aug 25	&	55432.554	&	12.06	&	13.46	&	14.60	&	16.23	&	1.3m & ANDOR\\
2010 Aug 26	&	55434.535	&	12.03	&	13.41	&	14.54	&	16.18	&	1.3m & ANDOR\\
2010 Aug 27	&	55435.551	&	12.00	&	13.37	&	14.50	&	16.13	&	1.3m & ANDOR\\
2010 Sep 08 & 55447.609 & 11.92 & 13.28 & 14.38 & -     & Sch  & FLI\\
2010 Sep 09 & 55448.561 & 11.89 & 13.25 & 14.34 & -     & Sch  & FLI\\
2010 Sep 10 & 55449.589 & 11.91 & 13.26 & 14.37 & -     & Sch  & FLI\\
2010 Sep 20	&	55459.581	&	11.87	&	13.21	&	14.32	&	15.93	&	1.3m & ANDOR\\
2010 Oct 12	&	55481.517	&	11.70	&	13.02	&	14.11	&	-	    &	1.3m & ANDOR\\
2010 Oct 29	&	55499.455	&	11.84	&	13.09	&	14.17	&	15.79	&	2m   & VA\\
2010 Oct 30	&	55500.483	&	11.82	&	13.08	&	14.16	&	15.78	&	2m   & VA\\
2010 Oct 31	&	55501.476	&	11.82	&	13.07	&	14.16	&	15.78	&	2m	 & VA\\
2010 Oct 31	&	55501.493	&	11.79	&	13.09	&	14.18	&	15.78	&	Sch	 & FLI\\
2010 Nov 01	&	55502.476	&	11.81	&	13.06	&	14.13	&	15.75	&	2m   & VA\\
2010 Nov 02	&	55502.522	&	11.78	&	13.09	&	14.17	&	15.73	&	Sch  & FLI\\
2010 Nov 02	&	55503.485	&	11.80	&	13.12	&	14.19	&	15.80	&	Sch  & FLI\\
2010 Nov 03	&	55504.458	&	11.80	&	13.11	&	14.20	&	15.75	&	Sch  & FLI\\
2010 Nov 04	&	55505.430	&	11.81	&	13.11	&	14.19	&	15.77	&	Sch  & FLI\\
2010 Nov 05	&	55506.313	&	11.82	&	13.13	&	14.20	&	15.77	&	Sch  & FLI\\
2010 Nov 06	&	55507.492	&	11.83	&	13.16	&	14.25	&	15.84	&	Sch  & FLI\\
2011 Jan 01	&	55563.377	&	11.80	&	13.11	&	14.17	&	15.76	&	Sch  & FLI\\
2011 Jan 06	&	55568.360	&	11.91	&	13.21	&	14.31	&	15.96	&	2m   & VA\\
2011 Jan 08	&	55570.257	&	11.94	&	13.23	&	14.34	&	15.99	&	2m   & VA\\
2011 Jan 09	&	55571.364	&	11.92	&	13.24	&	14.35	&	16.02	&	2m   & VA\\
2011 Jan 11	&	55573.289	&	11.94	&	13.25	&	14.37	&	16.05	&	2m   & VA\\
2011 Feb 06 & 55599.305 & 12.05 & 13.43 & 14.59 & 16.21 & Sch  & FLI\\
2011 Feb 07 & 55600.324 & 12.06 & 13.45 & 14.60 & 16.27 & Sch  & FLI\\
2011 Apr 04	&	55656.296	&	11.72	&	13.01	&	14.06	&	15.61	&	Sch  & FLI\\
2011 Apr 09	&	55661.280	&	11.71	&	12.99	&	-	    &	-   	&	2m   & VA\\
2011 Aug 17	&	55790.579	&	11.84	&	13.11	&	14.17	&	15.70	&	1.3m & ANDOR\\
2011 Aug 18	&	55791.577	&	11.83	&	13.10	&	14.15	&	15.68	&	1.3m & ANDOR\\
2011 Aug 24 & 55797.524 & 11.81 & 13.09 & 14.12 & 15.67 & Sch  & FLI\\
2011 Aug 25 & 55798.559 & 11.80 & 13.07 & 14.10 & 15.63 & Sch  & FLI\\
2011 Aug 26 & 55799.535 & 11.82 & 13.08 & 14.11 & 15.64 & Sch  & FLI\\
2011 Sep 11	&	55815.538	&	11.84	&	13.10	&	14.14	&	15.66	&	1.3m & ANDOR\\
2011 Sep 12	&	55816.550	&	11.82	&	13.08	&	14.12	&	15.64	&	1.3m & ANDOR\\
2011 Sep 19	&	55824.491	&	11.83	&	13.09	&	14.14	&	15.64	&	1.3m & ANDOR\\
2011 Oct 07	&	55842.444	&	11.92	&	13.21	&	14.27	&	15.81	&	1.3m & ANDOR\\
2011 Oct 13	&	55848.442	&	11.97	&	13.28	&	14.36	&	15.92	&	1.3m & ANDOR\\
2011 Oct 30	&	55865.451	&	12.06	&	13.44	&	14.53	&	16.16	&	2m   & VA\\
2011 Oct 31	&	55866.448	&	12.03	&	13.41	&	14.53	&	16.16	&	2m   & VA\\
2011 Nov 26	&	55892.390	&	12.24	&	13.64	&	14.76	&	16.45	&	2m   & VA\\
2011 Nov 27	&	55893.362	&	12.18	&	13.53	&	14.63	&	16.26	&	Sch	 & FLI\\
2011 Nov 28	&	55894.329	&	12.15	&	13.51	&	14.60	&	16.23	&	Sch	 & FLI\\
2011 Nov 29	&	55895.433	&	12.14	&	13.48	&	14.58	&	16.21	&	Sch	 & FLI\\
2011 Nov 30	&	55896.395	&	12.15	&	13.51	&	14.59	&	16.22	&	Sch  & FLI\\
2011 Dec 29	&	55925.433	&	12.10	&	13.45	&	14.52	&	16.15	&	Sch	 & FLI\\
2012 Jan 01	&	55928.303	&	12.07	&	13.40	&	14.47	&	16.07	&	Sch	 & FLI\\
2012 Jan 29	&	55956.251	&	12.48	&	13.90	&	15.05	&	-	    &	2m   & VA\\
2012 Mar 16	&	56003.346	&	14.22	&	15.70	&	16.73	&	18.30	&	Sch  & FLI\\
2012 Mar 20 & 56007.379 & 14.21 & 15.65 & 16.75 & 18.20 & 60cm & FLI\\
2012 Mar 22 & 56009.268 & 14.25 & 15.69 & 16.78 & 18.28 & 60cm & FLI\\
2012 Mar 23 & 56010.275 & 14.25 & 15.68 & 16.80 & 18.30 & 60cm & FLI\\
2012 Mar 28	&	56016.346	&	14.29	&	15.74	&	16.82	&	18.23	&	2m   & VA\\
2012 Apr 03 & 56021.398 & 14.32 & 15.75 & 16.76 & -     & 60cm & FLI\\
2012 Apr 10	&	56028.305	&	14.30	&	15.73	&	16.85	&	-	    &	Sch  & FLI\\
2012 Apr 12	&	56030.293	&	14.29	&	15.73	&	16.83	&	18.21	&	Sch  & FLI\\
2012 Jul 31	&	56139.590	&	12.61	&	14.13	&	15.40	&	17.17	&	1.3m & ANDOR\\
2012 Aug 02	&	56141.568	&	12.50	&	13.99	&	15.25	&	17.01	&	1.3m & ANDOR\\
2012 Aug 03	&	56142.569	&	12.47	&	13.95	&	15.20	&	16.96	&	1.3m & ANDOR\\
2012 Aug 04	&	56143.590	&	12.44	&	13.93	&	15.20	&	-	    &	1.3m & ANDOR\\
2012 Aug 11	&	56150.601	&	12.37	&	13.84	&	15.06	&	16.72	&	1.3m & ANDOR\\
2012 Aug 13	&	56152.609	&	12.38	&	13.84	&	15.06	&	16.73	&	1.3m & ANDOR\\
2012 Aug 14	&	56153.598	&	12.36	&	13.83	&	15.06	&	16.71	&	1.3m & ANDOR\\
2012 Aug 17	&	56156.619	&	12.40	&	13.87	&	15.09	&	16.81	&	1.3m & ANDOR\\
2012 Aug 18	&	56157.604	&	12.39	&	13.85	&	15.07	&	16.76	&	1.3m & ANDOR\\
2012 Aug 20 & 56159.545 & 12.42 & 13.88 & 15.08 & 16.79 & Sch  & FLI\\
2012 Aug 21	&	56160.556	&	12.41	&	13.87	&	15.10	&	16.81	&	1.3m & ANDOR\\
2012 Aug 21 & 56160.557 & 12.42 & 13.90 & 15.11 & 16.85 & Sch  & FLI\\
2012 Aug 22 & 56161.564 & 12.42 & 13.89 & 15.10 & 16.79 & Sch  & FLI\\
2012 Sep 03	&	56173.529	&	12.45	&	13.92	&	15.11	&	16.81	&	1.3m & ANDOR\\
2012 Sep 04	&	56174.531	&	12.45	&	13.93	&	15.13	&	16.82	&	1.3m & ANDOR\\
2012 Sep 05 & 56175.483 & 12.49 & 13.97 & 15.15 & 16.76 & Sch  & FLI\\
2012 Sep 06 & 56176.513 & 12.50 & 13.99 & 15.17 & 16.87 & Sch  & FLI\\
2012 Sep 09 & 56179.502 & 12.53 & 14.05 & 15.26 & 16.94 & 1.3m & ANDOR\\
2012 Sep 09 & 56180.387 & 12.61 & 14.11 & 15.32 & -     & 60cm & FLI\\
2012 Sep 10 & 56180.515 & 12.59 & 14.13 & 15.37 & 17.07 & 1.3m & ANDOR\\
2012 Sep 11 & 56182.476 & 12.65 & 14.20 & 15.45 & 17.16 & 1.3m & ANDOR\\
2012 Sep 13 & 56183.520 & 12.67 & 14.25 & 15.50 & 17.22 & 1.3m & ANDOR\\
2012 Sep 23 & 56193.515 & 12.68 & 14.29 & 15.58 & 17.32 & 1.3m & ANDOR\\
2012 Sep 23 & 56193.523 & 12.70 & 14.30 & 15.58 & 17.31 & Sch  & FLI\\
2012 Sep 24 & 56194.504 & 12.71 & 14.30 & 15.57 & 17.28 & Sch  & FLI\\
2012 Oct 09 & 56209.526 & 12.16 & 13.54 & 14.68 & 16.31 & Sch  & FLI\\
2012 Oct 10 & 56211.496 & 12.13 & 13.49 & 14.60 & 16.23 & Sch  & FLI\\
2012 Oct 11 & 56212.405 & 12.09 & 13.43 & 14.54 & 16.17 & 60cm & FLI\\
2012 Oct 13 & 56214.357 & 12.08 & 13.39 & 14.52 & 16.13 & 2m   & VA\\
2012 Oct 25 & 56226.402 & 12.03 & 13.37 & 14.45 & 16.04 & Sch  & FLI\\
2012 Oct 26 & 56227.480 & 12.04 & 13.40 & 14.50 & 16.12 & Sch  & FLI\\
2012 Nov 17 & 56249.408 & 12.07 & 13.42 & 14.55 & 16.21 & Sch  & FLI\\
2012 Nov 18 & 56250.435 & 12.06 & 13.42 & 14.53 & 16.17 & Sch  & FLI\\
2012 Nov 27 & 56258.648 & 12.04 & 13.38 & 14.47 & 16.07 & Sch  & FLI\\
2012 Dec 12 & 56274.270 & 12.08 & 13.45 & 14.51 & 16.11 & 2m   & VA\\
2012 Dec 14 & 56275.532 & 12.06 & 13.44 & 14.49 & 16.11 & 2m   & VA\\
2012 Dec 14 & 56276.393 & 12.08 & 13.44 & 14.49 & 16.10 & 2m   & VA\\
2012 Dec 30 & 56292.345 & 12.04 & 13.38 & 14.42 & 16.17 & 60cm & FLI\\
2013 Jan 01 & 56294.359 & 12.03 & 13.34 & 14.41 & 16.07 & 60cm & FLI\\
2013 Jan 03 & 56296.382 & 12.02 & 13.34 & 14.42 & 16.03 & 60cm & FLI\\
2013 Jan 16 & 56309.306 & 12.03 & 13.33 & 14.41 & 16.03 & Sch  & FLI\\
2013 Jan 19 & 56312.308 & 11.99 & 13.25 & 14.35 & 15.97 & 2m   & VA\\
2013 Feb 02 & 56326.247 & 11.91 & 13.20 & 14.23 & 15.81 & Sch  & FLI\\
2013 Feb 04 & 56328.416 & 11.90 & 13.18 & 14.21 & 15.75 & Sch  & FLI\\
2013 Feb 05 & 56329.401 & 11.91 & 13.19 & 14.24 & 15.80 & Sch  & FLI\\
2013 Mar 04 & 56356.461 & -     & 13.16 & 14.24 & -     & 60cm & FLI\\
2013 Mar 05 & 56357.441 & 11.91 & 13.19 & 14.24 & 15.85 & 60cm & FLI\\
2013 Mar 17 & 56369.361 & 11.92 & 13.23 & 14.24 & 15.79 & 2m   & VA\\
2013 Mar 19 & 56371.361 & 11.97 & 13.27 & 14.28 & 15.80 & 2m   & VA\\
2013 Apr 10 & 56393.316 & 11.96 & 13.26 & 14.30 & 15.85 & Sch  & FLI\\
2013 Apr 11 & 56394.305 & 11.96 & 13.27 & 14.31 & 15.87 & Sch  & FLI\\
2013 Apr 12 & 56395.289 & 11.96 & 13.26 & -     & -     & Sch  & FLI\\
\end{longtable}
}

\subsection{Spectral observations}

At the time of our photometric monitoring of V582 Aur, a total of sixteen optical spectra of the star were obtained.
High-, medium-, and low-resolution spectroscopy of V582 Aur was performed using spectral equipment in three observatories: Asiago (Italy), Haute-Provence (France), and Skinakas (Greece).
All data reduction was performed within IRAF, Table 3 provides a log of spectral observations.

Low-dispersion, absolutely fluxed spectra of V582 Aur were obtained with the Asiago 1.22 m + B\&C telescope operated by the Department of Physics
and Astronomy of the University of Padova.  
A 300 ln/mm grating blazed at 5000 \AA\ provided a dispersion of 2.31 \AA/pix and a FWHM (PSF) $\sim$3.0 pix.  
The detector was an Andor iDus DU440 CCD camera with a 2048$\times$512 pixel array, which is of high UV efficiency as the whole spectrograph optical train.
The primary spectrophotometric standard star was HR 1729, which was only a few degrees away and observed immediately before or after V582 Aur.  
High-resolution spectra of V582 Aur were secured with the Asiago 1.82 m telescope equipped with an REOSC echelle spectrograph and an Andor DW436BV CCD camera, housing a back-illuminated E2V CCD4240 AIMO detector with a 2048$\times$2048 pixel array.  
A binning of 2$\times$2 provided a resolving power of 12000.  
At both telescopes the slit was oriented along the parallactic angle and widened to 2.0 arcsec.  

One high-resolution and one low-resolution spectrum of V582 Aur were obtained with the 1.93 m telescope at the Haute-Provence Observatory. 
The high-resolution spectrum was obtained with the cross-dispersed echelle spectrograph (Sophie) on 2010 January 15 and the low-resolution spectrum with the long-slit Cassegrain spectrograph (Carelec) on 2012 January 18.
The CCD chips used are EEV 42-20 CCD ($2048\times1024$ pixels) for Carelec and EEV ($4096\times2048$) for Sophie.
 
Observations in the Skinakas Observatory were carried out with the focal reducer of the 1.3 m RC telescope and ISA 608 spectral CCD camera ($2000\times800$ pixels, 15$\times15$ $\mu$m) on 2012 August 31, September 1, and September 23.
Two gratings (1300 and 2400 lines per mm) and 160$\mu$m slit were used.
The first grating yield a resolving power $\lambda/\Delta\lambda$ $\sim$ 1300, while the second grating yielded $\lambda/\Delta\lambda$ $\sim$ 2500 at H$\alpha$ line.
The exposures of V582 Aur were followed immediately by an exposure of an FeHeNeAr comparison lamp and exposure of a spectrophotometric standard star.

\begin{table*}
\caption{Journal of spectroscopic observations}
\label{table:3}
\centering
\begin{tabular}{lrrlll}
\hline
\hline
\noalign{\smallskip}
Date & \multicolumn{1}{c}{UT} & \multicolumn{1}{c}{Exp. t.} & Dispersion or& $\lambda$ range& Tel.\\
 & & (sec) & resolving power & (\AA) & \\
\noalign{\smallskip}
\hline
\noalign{\smallskip}
2009 Aug 06 & 02:34  & 3600  & disp. 2.31 \AA/pix   & 4000$-$6950   &  1.22m+B\&C \\
2010 Jan 15 & 23:26  & 3000  & res. pow. 45 000     & 3872$-$6943   &  1.93m+SOPHIE\\
2011 Dec 21 & 22:04  & 2700  & disp. 2.31 \AA/pix   & 3700$-$7550   &  1.22m+B\&C \\  
2012 Jan 13 & 19:27  & 3600  & res. pow. 12 000     & 4400$-$7335   &  1.82m+REOSC\\
2012 Jan 18 & 20:22  & 2400  & res. pow. 900        & 3600$-$7000   &  1.93m+CARELEC\\
2012 Feb 08 & 21:38  & 900   & res. pow. 12 000     & 4400$-$7335   &  1.82m+REOSC\\  
2012 Feb 18 & 20:02  & 900   & disp. 2.31 \AA/pix   & 3700$-$7550   &  1.22m+B\&C\\    
2012 Mar 02 & 20:48  & 1800  & res. pow. 12 000     & 4400$-$7335   &  1.82m+REOSC\\  
2012 Mar 30 & 19:01  & 1800  & disp. 2.31 \AA/pix   & 3700$-$7550   &  1.22m+B\&C\\           
2012 Aug 31 & 01:41  & 3600  & disp. 0.48 \AA/pix   & 5750$-$6720   &  1.30m+focal red.\\
2012 Sep 01 & 23:50  & 1800  & disp. 1.05 \AA/pix   & 5490$-$7590   &  1.30m+focal red.\\
2012 Sep 23 & 00:22  & 3600  & disp. 1.05 \AA/pix   & 5490$-$7590   &  1.30m+focal red.\\
2012 Oct 21 & 22:55  & 900   & disp. 2.31 \AA/pix   & 3700$-$7550   &  1.22m+B\&C   \\
2012 Oct 29 & 23:08  & 900   & res. pow. 12 000     & 4400$-$7335   &  1.82m+REOSC\\  
2012 Nov 15 & 22:02  & 1200  & disp. 0.61 \AA/pix   & 5680$-$6910   &  1.22m+B\&C \\
2012 Dec 28 & 20:15  & 1500  & res. pow. 12 000     & 4400$-$7335   &  1.82m+REOSC\\  
\hline
\end{tabular}
\end{table*}

\subsection{Archival photographic observations}

The construction of the historical light curves of FUors could be very important for determining the exact moment of the beginning of the outburst and the time to reach the maximum light.
Another important option is to study the pre-outburst variability of the star.
The only possibility for such a study is a search in the photographic plate archives at the astronomical observatories around the world.
Most suitable for this purpose are the plate archives of the big Schmidt telescopes that have a large field of view.
Unfortunately, the collection and analysis of old photographic observations requires a very long and laborious amount of work.
In this paper, we present our first result of exploring the whole photographic plate stack of the 105/150 cm Schmidt telescope at Kiso Observatory (Japan), and several plates around the expected time of the outburst obtained with the 67/92 cm Schmidt telescope at Asiago Observatory (Italy).
We also used the digitized plates from the Palomar Schmidt telescope, available via the website of the Space Telescope Science Institute.
In addition we checked for archival observations of V582 Aur in the photographic plate collection of the 2 m RCC and the 50/70 cm Schmidt telescopes at NAO Rozhen (Bulgaria), but found none.

The plates from Asiago Schmidt telescope are inspected visually through a high-quality Carl Zeiss microscope offering a variety of magnifications (Munari et al. 2001). 
The magnitude is then derived by comparing the stars in the photometric sequence with the variable, identifying those that are more closely bracketing the variable.
If ``a" and ``b" are such two stars of the sequence, visual inspection estimate the quantities n1, n2, which represent the fraction of the total 
a-b magnitude difference by which the variable V is fainter than ``a" and brighter than ``b":  a - n1 - V - n2 - b. 
The magnitude of V follows from a simple proportion. If more than one such pair is available, more estimates are derived and weighted according to the ``a-b", ``c-d" etc. mag interval.
Typical estimated errors are of the order of 0.10 mag.

The plates from Kiso Schmidt telescope were scanned with Canon CanoScan LiDE 600F portable scanner, which has 1200 dpi resolution.
Each photographic plate was put on the scanner glass plate, and three sheets of white paper were stacked on the photographic plate.
A fluorescent tube was used to light the photographic plates.

Aperture photometry of the digitized plates was performed with DAOPHOT routines using the same aperture radius and the background annulus as for CCD photometry. 
The results of the measured magnitudes of V582 Aur from the archival photographic plates are given in Table 4. 
The columns provide the name of the observatory, the plate number, date and Julian date (J.D.) of observation, photographic emulsions and filters used, the magnitude estimated or plate limit, and the corresponding errors.

\begin{table*}
\centering
\caption[]{Photometric data from the photographic observations of V582 Aur}
\begin{tabular}{llllllll}
              \hline
             \hline
\noalign {\smallskip}  
Observatory	&	Plate No.	&	Date	&	J.D.	&	Emulsion	&	Filter	&	Magnitude	&	Er.	\\
            \hline
\noalign {\smallskip} 
Palomar	&	001315E	&	1954 Dec 29	&	2435105.778	&	103aE &	Plexi	&	R=16.87  &	$\pm$0.06	\\
Palomar	&	001315O	&	1954 Dec 29	&	2435105.809	&	103aO	&	none	&	pg=19.5	 &	$\pm$0.1	\\
Kiso    & 000901  & 1977 Oct 10 & 2443427.258 & IN    & RG695 & I$>$15.0 &    \\
Kiso    & 001521  & 1978 Mar 20 & 2443587.927 & IN    & RG695 & I=15.8   &  $\pm$0.2  \\
Kiso    & 001965  & 1979 Jan 02 & 2443875.957 & 103aE & RG645 & R=16.6   &  $\pm$0.2  \\
Kiso    & 001967  & 1979 Jan 02 & 2443876.159 & IIaD  & GG495 & V=18.0   &  $\pm$0.3  \\
Kiso    & 001973  & 1979 Jan 04 & 2443878.072 & 103aE & RG645 & R=17.1   &  $\pm$0.2  \\
Kiso    & 002409  & 1979 Nov 15 & 2444193.160 & IIaD  & GG495 & V$>$17.6 &    \\
Kiso    & 002413  & 1979 Nov 15 & 2444193.287 & IN    & RG695 & I$>$15.0 &    \\
Kiso    & 002492  & 1979 Dec 14 & 2444222.201 & 103aE & RG645 & R=17.2   &  $\pm$0.2  \\
Kiso    & 002529  & 1979 Dec 22 & 2444230.004 & 103aE & RG645 & R$>$16.7 &    \\
Kiso    & 003010  & 1980 Nov 14 & 2444558.256 & IIaD  & GG495 & V$>$17.5 &    \\
Kiso    & 003057  & 1980 Dec 10 & 2444584.059 & IIaD  & GG495 & V$>$17.5 &    \\
Palomar	&	000310V	&	1982 Oct 21	&	2445263.955	&	IIaD	&	W12	  &	V=18.9	 &	$\pm$0.1	\\
Asiago	&	011872	&	1983 Jan 12	&	2445347.360	&	103aO	&	GG13	&	B$>$18.5 &		\\
Asiago	&	011873	&	1983 Jan 12	&	2445583.380	&	IN	  &	RG5 	&	I$>$15.9 &		\\
Asiago	&	012690	&	1984 Nov 29	&	2446036.444	&	IN  	&	RG5 	&	I$>$15.9 &		\\
Asiago	&	013276	&	1986 Jan 17	&	2446448.436	&	IN  	&	RG5 	&	I=15.25  &	$\pm$0.08	\\
Palomar	&	001013 	&	1986 Dec 29	&	2446793.756	&	IIIaJ	&	GG385	&	B=15.74  &	$\pm$0.05	\\
Asiago	&	013706	&	1987 Feb 01	&	2446828.377	&	103aD	&	GG14 	&	V=14.25  &	$\pm$0.08	\\
Palomar	&	002236  &	1988 Dec 01	&	2447496.817	&	IIIaF	&	RG610	&	R=13.85  &	$\pm$0.05	\\
Palomar	&	002812  &	1989 Oct 04	&	2447803.958	&	IIIaF	&	RG610	&	R=13.86  &	$\pm$0.06	\\
Palomar	&	002935	&	1989 Nov 19	&	2447849.846	&	IVN	  &	RG9	  &	I=12.21  &	$\pm$0.03	\\
Palomar	&	005536  &	1993 Oct 23	&	2449283.920	&	IIIaJ	&	GG385	&	B=15.86  &	$\pm$0.06	\\
Palomar	&	007538	&	1997 Oct 30	&	2450751.851	&	IVN	  &	RG9	  &	I=12.06  &	$\pm$0.03	\\
\hline
\end{tabular}
\end{table*}

\section{Results}
\subsection{Photometric monitoring}

The historical $BVRI$ light curves of V582 Aur from all available photometric observations are plotted in Fig. 3. 
On the figure, the filled diamonds represent the CCD observations from the present paper,
the filled circles photographic data from the 67/92 cm Asiago Schmidt telescope,
the filled triangles photographic data from the 105/150 cm Kiso Schmidt telescope,
and the filled squares photographic data from the Oschin Schmidt Telescope on Palomar.

\begin{figure*}
   \centering
   \includegraphics[width=18cm]{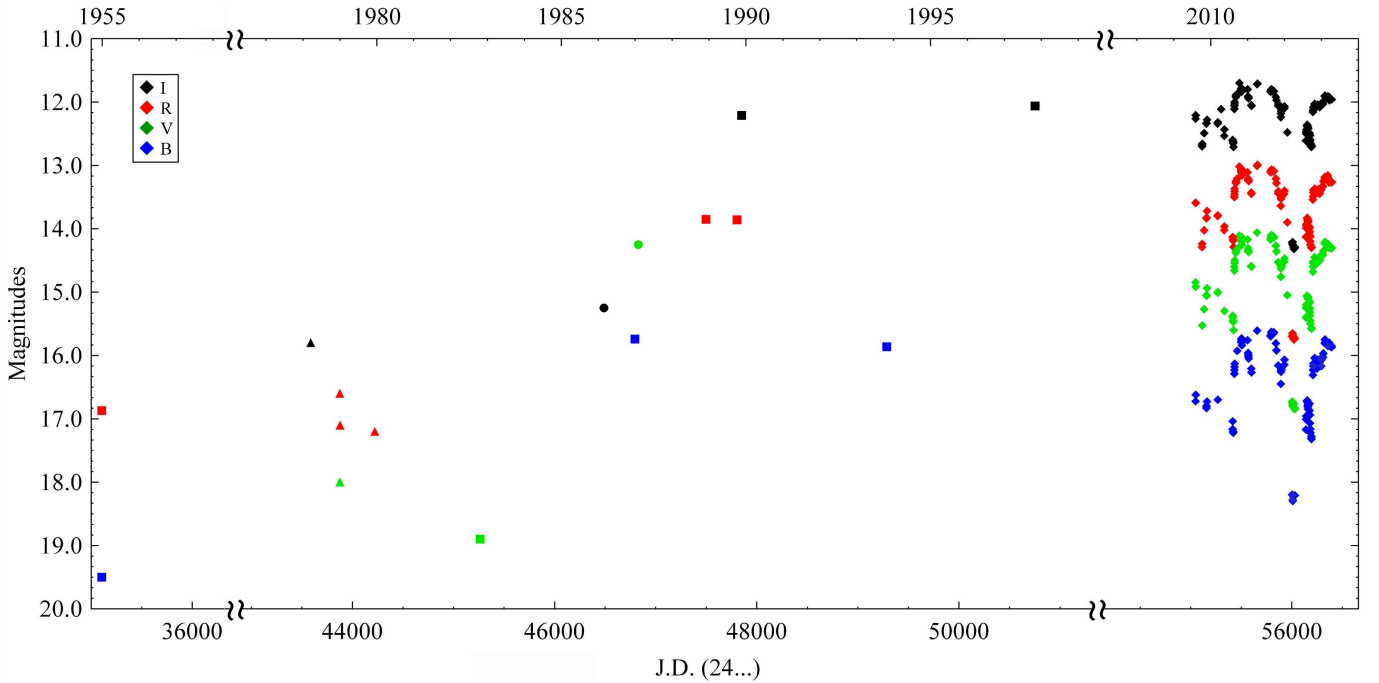}
      \caption{Historical $BVRI$ light curves of V582 Aur for the period 1954 December $-$ 2013 February.}
         \label{Fig3}
  \end{figure*}

The light curve of V582 Aur allows us to conclude that its photometric behavior is similar to that of classical FUor stars.
The $R$ and $V$ magnitudes measured from Kiso and Palomar photographic plates suggest that before the outburst the star was variable with an amplitude at about 1 mag. 
Archival photographic observations indicate that the increase in brightness began in late 1984 or early 1985 and the star brightness reached its maximum value in 1986 January.
Hence, the rise in brightness is relatively fast at about one year and the registered amplitude is $\sim$3.6 mag. ($V$).
The photographic data from Palomar plates indicate that during the first decade after the outburst (1986-1997) the star keeps its maximum brightness with approximately constant $B$, $R$, and $I$ magnitudes (Table 4).
The CCD photometric data reported in the present paper show a very strong and fast variability in brightness, while the star remains in a state close to the maximum brightness most of the time. 
Consequently, the outburst of V582 Aur continued for approximately 28 years. 

Figure 4 shows the $R$-light curve of V582 Aur for the period of our CCD observations and the dates of spectroscopic observations (marked by blue arrows).
The present photometric data show large amplitude variations ($\Delta R\sim2\fm8$) during the 3.5 year period of observations. 
The photometric variability does not show periodicity, as the brightness variations sometimes occur very rapidly in the time scale of days and weeks.
The fastest changes in brightness were registered in 2010 August, when for a twelve-day period the stellar magnitude increased by $1\fm10$ ($V$) and in 2012 September-October, when for a seventeen-day period the stellar magnitude increased by $1\fm05$ ($V$). 
But the most remarkable event in the light curve of V582 Aur is the large drop in brightness during the spring of 2012.
A strong decline in brightness with $2\fm26$ ($V$) from January 1 to March 16 was observed and the star remained in a state of very low brightness for at least one month. 
The observed $BVRI$ magnitudes during this period (2012 March-April) are only by 1-1.5 mag. higher than the pre-outburst magnitudes measured from the archival photographic plates.

\begin{figure}
   \centering
   \includegraphics[width=9cm]{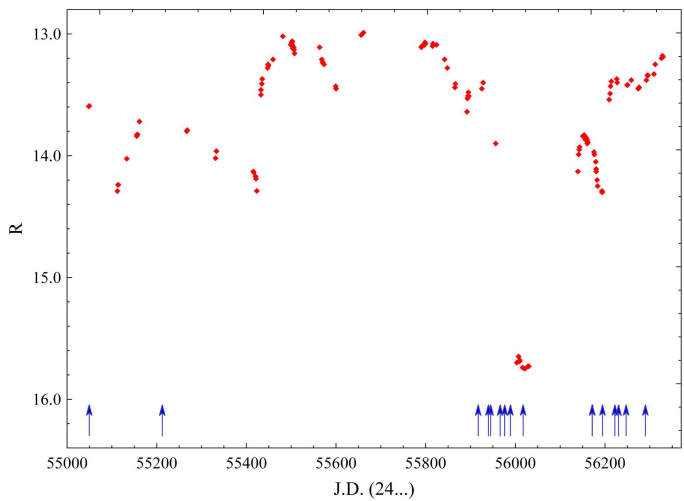}
   \caption{$R$-light curve of V582 Aur for the period 2009 August $-$ 2013 February. Dates of spectroscopic observations are marked by arrows.}
    \end{figure}

Another important result from our photometric study is the variation of color indices with stellar brightness.
Figure 5 shows the measured color indices $V-I$ versus stellar magnitude $V$ during the period of our observations.
A clear dependence can be seen over almost the entire period: the star becomes redder as it fades.
Such color variations are typical for FUor stars, which have a relatively fast set in brightness, as for example, V1057 Cyg (Hartmann \& Kenyon 1996).
In the case of V582 Aur, the changes of color indices occur very rapidly within days and weeks as we observe drops in brightness accompanied by reddening and increases in brightness accompanied by blue color indices.
But the multicolor photometric data obtained through the deep minimum in brightness (spring 2012) indicated a different relationship.
From a certain turning point, V582 Aur gets bluer, fading further to $V$ $\sim16\fm8$ on 2012 March. 
Such color variations are observed as well for the $V-R$ and $B-V$ indices.
The observed change of color indices suggests the existence of a color reversal (or so-called ``blueing") in the minimum light, a typical feature of PMS stars from the UXor type.

\begin{figure}
   \centering
   \includegraphics[width=9cm]{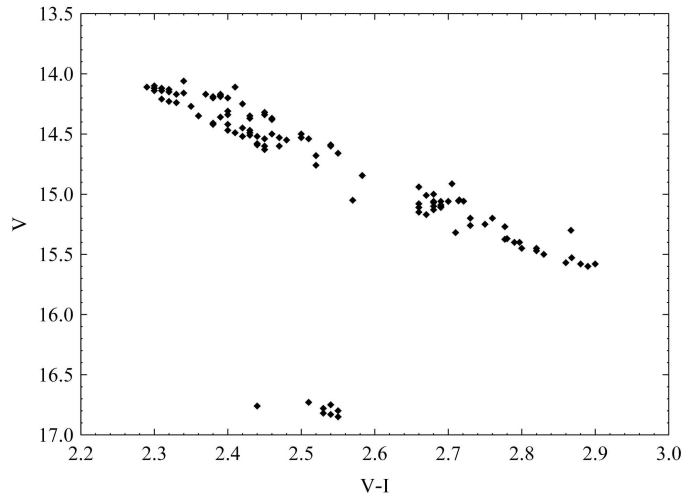}
   \caption{$V-I$ vs. $V$ color-magnitude diagram for V582 Aur during the outburst (2009 August - 2013 February).}
    \end{figure}

\subsection{Spectroscopic monitoring}

Besides the strong photometric variability, relatively rapid changes of the profiles and intensities at different spectral lines have been registered.
The first low-dispersion spectrum obtained on 2009 August 6 is dominated by absorption lines of the Balmer series (H$\alpha$ and H$\beta$), Na I D, and Ba II 6497 $\AA$, and emission lines are not noticed. 
The spectrum appears like a supergiant of temperature about 5500 K, but the Li I 6707 $\AA$ line is weak, not emerging above the noise (Munari et al. 2009).
On the high-resolution spectrum obtained on 2010 January 15, the H$\alpha$ line and Na I doublet show the P Cyg profiles, which are typical of FUor stars (Fig. 6). 
The broad blueshifted H$\alpha$ absorption seems to be saturated, extending to about -800 km/s due to powerful, rapidly expanding wind.
All high- and medium-resolution spectra, except those obtained in 2012 February-March, show the same spectral features, but with changes in the P Cyg profiles of H$\alpha$ line (Fig. 6).
Moreover, at increased brightness of the star, the absorption component becomes deeper and wider and the emission component becomes less intensive.
Therefore, there is a noticeable correlation between the variations in the optical depth of H$\alpha$ line and the photometric properties of the star. 

\begin{figure*}
   \centering
   \includegraphics[width=5.5cm]{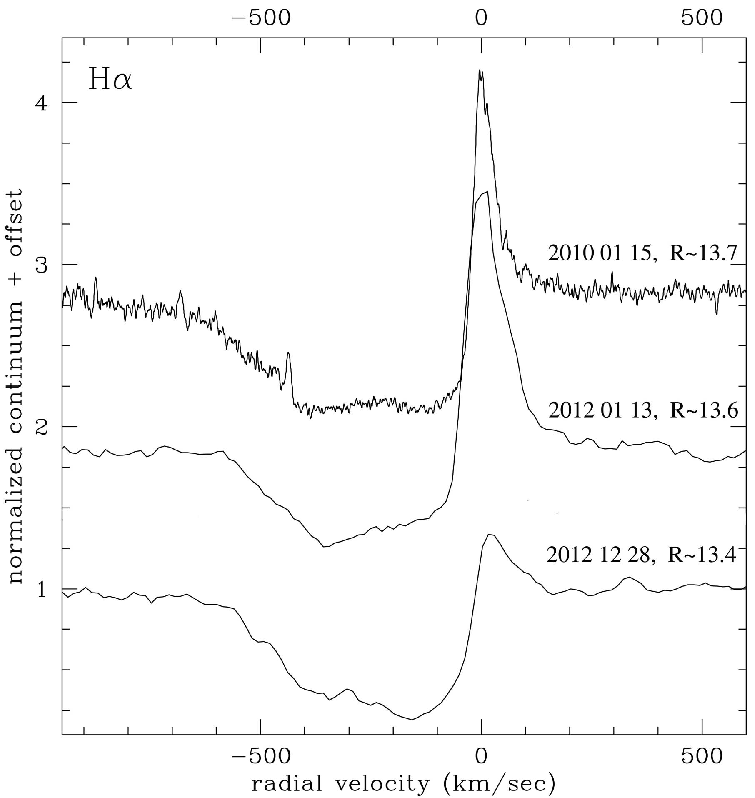}
   \includegraphics[width=5.6cm]{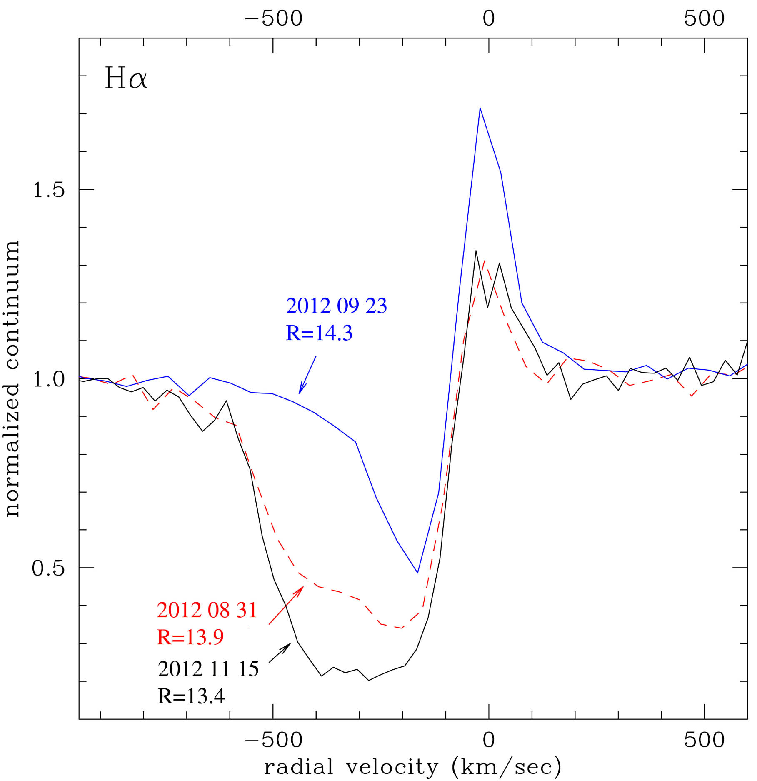}
   \includegraphics[width=6cm]{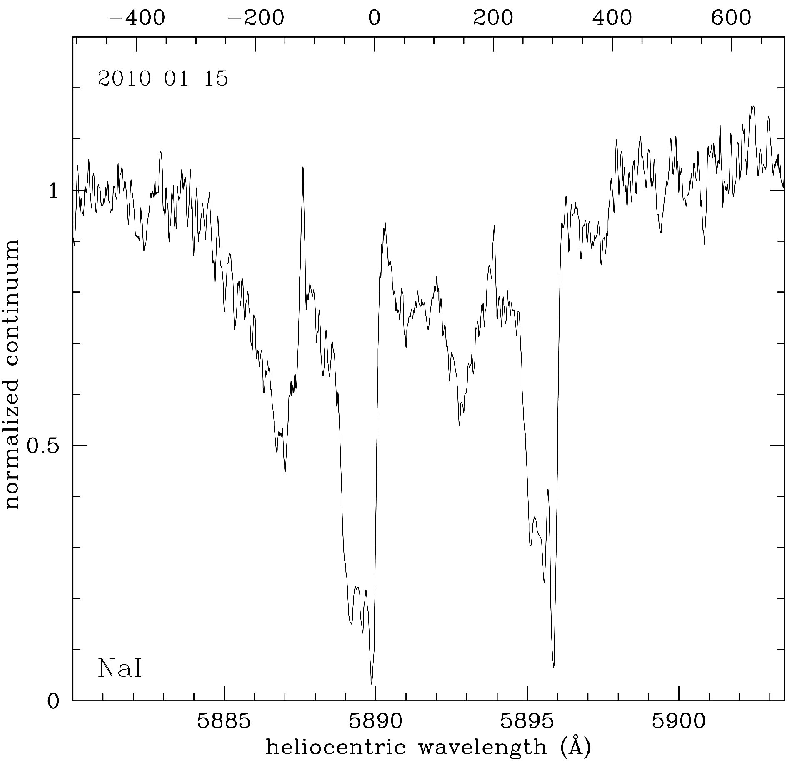}
   \caption{Profiles of H$\alpha$ and Na I D lines from different periods of observations. 
   {\it Left}: high-resolution profiles of H$\alpha$ obtained with the SOPHIE and REOSC echelle spectrograph.
   {\it Center}: medium-resolution profiles of H$\alpha$ from the Skinakas and Asiago telescopes. 
   {\it Right}: high-resolution profile of Na I D obtained with the SOPHIE echelle spectrograph.}
    \end{figure*}

Figure 7 presents a comparison between the low-resolution spectra of V582 Aur, which are obtained before, during, and after the large drop in brightness in spring 2012.
During this event, the spectroscopic properties of the star changed dramatically.
The absorption lines and P Cyg profiles disappeared from the spectrum, and only the H$\alpha$ emission line without an absorption component remained.
Therefore, during the decline in brightness with $\sim$ $2\fm26$ ($V$), the spectral features of the stellar spectrum change from a typical FUor to typical T Tauri star spectrum.
With the ending of the deep decline in brightness of V582 Aur, the spectral characteristics change significantly again. 
The absorption lines of H$\alpha$, H$\beta$, Na I D, and Ba II 6497 $\AA$, along with the P Cyg profiles at H$\alpha$ and Na I D, appear again in 2012 August - September. 
A significant change in the SED of V582 Aur relative to spectra obtained at periods of high and low brightness is also observed.

\begin{figure*}
   \centering
   \includegraphics[width=16cm]{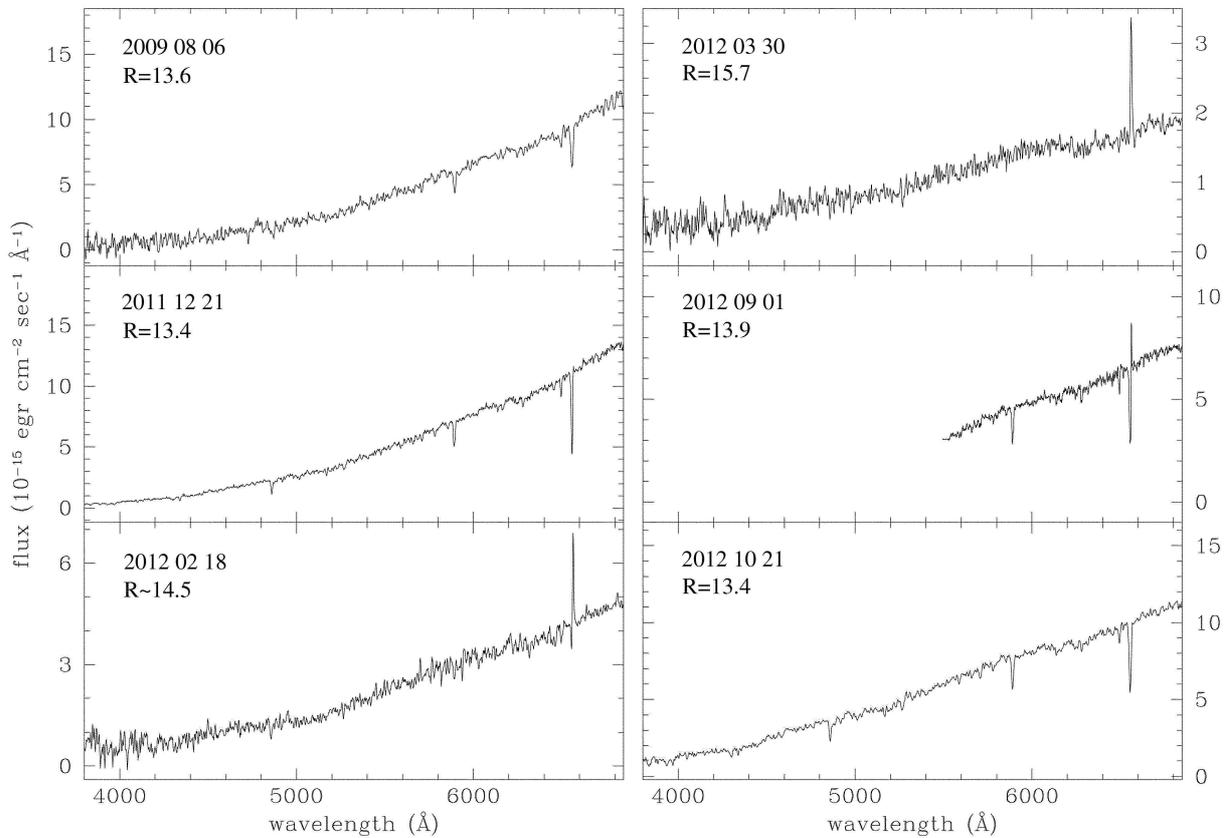}
   \caption{Low-resolution spectra of V582 Aur.}
    \end{figure*}

To classify the absorption spectrum of V582 Aur, we compared our spectra with the Asiago atlas of MKK spectral types observed with exactly the same    instrumental configuration of V582 Aur (Munari 2013, in preparation).
We took the absolutely fluxed spectra of V582 Aur and the MKK atlas and continuum normalized them using the same function (a Legendre polynomial of fifth order limited to the range of wavelength covered in Figure 8, which corresponds to those recommended for the classification within the MKK system). 
As a first classification pass, we applied a simple $\chi^2$ matching to determine the area of the Herzsprung-Russell diagram on which our deeper analysis was focused. 
The match found by the $\chi^2$ is not perfect, since the stellar spectra originates in stationary atmospheres where a three-dimensional treatment is generally unnecessary, while the absorption lines in V582 Aur instead form in a moving medium, the wind. 
We then proceeded to refine the classification by using an eye inspection of the spectra and found that the closest (even though imperfect) match was for a G0I type star. 
Figure 8 shows how the properties of the absorption spectrum of V582 Aur are similar to those of supergiants of the G0 types.
 
\begin{figure*}
   \centering
   \includegraphics[width=16cm]{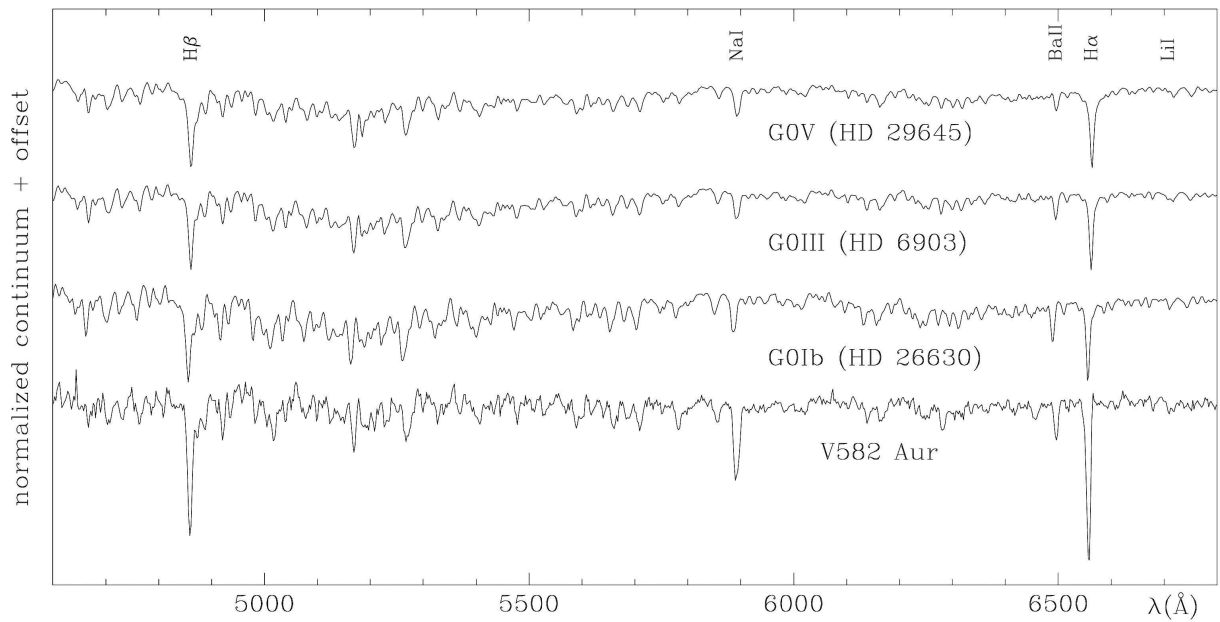}
   \caption{Low-resolution spectrum of V582 Aur obtained on 2012 December 21 is compared with spectra of the G supergiant stars HD 29645, HD 6903, and HD 26630 (from the Asiago spectral database).}
    \end{figure*}     

\section{Discussion}

Any case in which a large amplitude outburst of PMS star is observed gives rise to the question: FUor or EXor?
The main differences between these two types are the spectral appearance (presence or absence of certain spectral lines, their profiles, and intensity) and the photometric properties (duration and amplitude of the outburst and the shape of light curves).
The outburst of V582 Aur differs from EXor eruptions both in duration and spectral appearance. 
Photometric and spectroscopic data so far accumulated imply that V582 Aur is indeed proceeding through a {\it bona fide} FUor outburst.  
The presence of a 3.6 mag. amplitude outburst in optic and a rise in brightness in the period 1984-1986 are well documented.
Despite the sparse and random photometry available during 1986 to 2009, all available data suggest that the star keeps its maximal brightness at this period.
Therefore, a large amplitude outburst with duration of several decades is registered.

The spectrum dominated with prominent absorption lines of Balmer series, Na I and Ba II, the strong P Cyg profiles at H$\alpha$ and Na I D, and the presence of a reflection nebula around the star are all well-established characteristics of the classical FUor objects.  
The only discrepancies are the relatively weak absorption line of LiI~6707~\AA\ and the very strong and fast photometric variability, which are not typical for classical FUor stars. 
The observed profile of the H$\alpha$ line is highly variable, with a deep and high-velocity blueshifted absorption component. 
This feature can be interpreted as evidence of a strong and time-variable outflow driven by the central FUor object. 

The strong photometric variability observed over the 3.5 year period can be explained by 1) time-variable extinction or 2) changes in accretion rate from the circumstellar disk onto the stellar surface. 
The variable accretion affects the speed and intensity of the stellar wind and is manifested as changes in the P Cyg profile of the H$\alpha$ line.
Similar transformation of several spectral lines from absorption to emission have been observed in the deep minima of some UXor stars (Rodgers et al. 2002, Eaton \& Herbst 1995).
The spectral changes were explained by the different sizes of the star and the circumstellar envelope: the occulting screen is larger than the star but the circumstellar envelope is significantly larger than the screen.
In contrast to FUor stars the absorption and emission components in this case have symmetrical profiles and P Cyg profiles are not observed.

During the large drop in brightness in the spring of 2012, an appearance of dust particles in the immediate circumstellar environment of the star was registered.
An evidence of this statement is the observed effect of color reversal on the color/magnitude diagram (Fig. 5). 
The widely accepted explanation of the color reversal effect is variations of the column density of dust in the line of sight to the star.
Normally the star becomes redder when its light is covered by dust clumps or filaments on the line of sight.
But when the obscuration rises sufficiently, the part of the scattered light in the total observed light become significant and the star color gets bluer. 
Then, the decline in brightness was caused by either the reformation of circumstellar dust or motion of dust clumps into the line of sight toward the star.
In the fall of 2012, the dust particles were sublimated by the stellar wind or removed from the line of sight and the star returned to its previous photometric condition.

The strong photometric variability at maximum light is typical for EXors, but not for FUors.
However, for several FUor objects similar short-time drops in brightness were registered. 
The well-known one is the 1980 minimum in the light curve of V1515 Cyg, a strong decrease in brightness by about $1\fm5$ (B) in few months (Kolotilov \& Petrov 1983). 
This minimum in brightness was explained by an obscuration from a dust material ejected from the star (Kenyon et al. 1991). 
Evidence for strong light variability in the time of set in brightness ($\Delta$$V$=1$\fm$2) during the period from 1986 to 1992 was reported in the photometric study of another FUor object, V1735 Cyg (Peneva et al. 2009).
A short drop in brightness was observed in 2009 (decrease by 0$\fm$4 ($I$) and return to its previous level) in the light curve of V733 Cep, which is  also recognized as a FUor object (Peneva et al. 2010).
Unfortunately, spectral observations during minimum light are not available for these objects, and the cause of such short-time drops in brightness is not known.

The large amplitude variability may result from the superposition of both phenomena, the variable accretion rate and time variable extinction, and it is very difficult to distinguish the two phenomena using only photometric data (Semkov \& Peneva 2012).
In recent studies, such a scenario is used to explain the light variability of two PMS objects with characteristics similar to those of FUor and EXor -- V1647 Ori (Aspin et al. 2009b, Aspin 2011) and V2492 Cyg (Hillenbrand et al. 2013, K{\'o}sp{\'a}l et al. 2013). 
It seems that the time variable extinction is characteristic not only of some Herbig Ae/Be stars (UXor variables) but is also a common phenomenon during the evolution of all types of PMS stars. 
In the case of V582 Aur, we have direct evidence from multicolor photometry, which suggests the presence of dust around the star during the decline in brightness.

Even though many PMS stars indicate evidence of time variable accretion, the physical cause of this phenomenon is still under discussion.
One of the possible reasons for the variable accretion rate could be fragmentation of the circumstellar disk.
Because the FUor phenomenon is probably repeatable, up to 50\% of the protostellar mass can be accumulated as a result of such episodes of strong accretion burst.
Stamatellos et al. (2011, 2012) suggest that episodic accretion may initially promote disc fragmentation. 
In the early stages of PMS evolution, fragmentation does not happen and disk accretion is assumed to be constant.
After several episodic accretion bursts, the circumstellar disk is gradually fragmented and thus prevents new FUor events.
Therefore, it can be supposed that FUor outbursts during different periods of stellar evolution may vary in amplitude, duration, and shape of the light curve due to the different state of disk fragmentation.
Strong accretion bursts may also be the triggering mechanism of planet formation with different masses inside the circumstellar disk.
If the above suggestions are correct, the studies of FUor and EXor objects would be useful not only for understanding stellar evolution but also for understanding the formation of planets and asteroids and the frequency of planetary systems.

\section{Conclusions}

Photometric data presented in this paper show the usefulness of systematically spectral and photometric monitoring of the regions of star formation.
These data can be used to detect new FUor or EXor events and to determine the type of the registered outbursts.

On the basis of our photometric monitoring over the past 3.5 years and the spectral properties at maximal light (a G0I supergiant spectrum with strong P Cyg profiles of H$\alpha$ and Na I D lines), we have confirmed that the observed outburst of V582 Aur is of the FUor type.
On the other hand, the observed effect of color reversal at the minimum light is evidence of the possible symbiotic nature of V582 Aur (FUor + UXor).
Therefore, the collection of new photometric data (from photographic plate archives and ongoing photometric monitoring) will be of great importance for a precise determination of the type of variability.

At the same time, according to existing observations the light curve of V582 Aur remains unique, confirming the hypothesis that each known FUor has a different rate of increase and decrease in brightness and a different light curve shape.
We plan to continue our spectroscopic and photometric monitoring of the star during the next few months and years and strongly encourage similar follow-up observations.

\begin{acknowledgements}
     This work was partly supported by grants DO 02-85 and DO 02-362 of the National Science Fund of the Ministry of Education, Youth and Science, Bulgaria. 
     The authors thank the Director of Skinakas Observatory Prof. I. Papamastorakis and Prof. I. Papadakis for granting telescope time. 
     We also thank Prof. R. Zamanov for the fruitful discussion on the obtained results.
     The Digitized Sky Survey was produced at the Space Telescope Science Institute under U.S. Government grant NAG W-2166. 
     The images of these surveys are based on photographic data obtained using the Oschin Schmidt Telescope on Palomar Mountain and the UK Schmidt Telescope. 
     The plates were processed into the present compressed digital form with the permission of these institutions. 
     This research has made use of the NASA Astrophysics Data System.
\end{acknowledgements}

\end{document}